\documentclass[pamm,a4paper,fleqn,oneside]{w-art}
\usepackage{times}
\usepackage{w-thm}
\theoremstyle{plain}

\theoremstyle{definition}

\usepackage[]{graphicx}
\chardef\bslash=`\\ 

\begin{document}

\title{Vibrations of a Pendulum with Oscillating Support and Extra Torque}


\author{Marek Borowiec\footnote{Corresponding
     author: e-mail: {\sf m.borowiec@pollub.pl}, Phone: +48\,500\,274\,146,
     Fax: +48\,815\,241\,004}\inst{1}} \address[\inst{1}]{Department of Applied 
Mechanics, Technical University of
Lublin, \\
Nadbystrzycka 36, PL-20-618 Lublin, Poland}
\author{Grzegorz Litak\footnote{g.litak@pollub.pl}\inst{1}}
\address[\inst{2}]{Institut f\"{u}r Mechanik und Mechatronik, Technische 
Universit\"{a}t
Wien, \\
Wiedner
Hauptstra$\beta$e
8--10, A-1040 Wien, Austria \\}
\author{Hans Troger\footnote{htroger@mch2ws4.mechanik.tuwien.ac.at}\inst{2}}
\begin{abstract}

The motion 
of a driven   planar pendulum with vertically periodically oscillating point of
suspension and under the action of an additional constant
torque is investigated.
We study the influence of the torque strength on the
transition to chaotic motions of the pendulum using Melnikov's analysis.
\end{abstract}
\maketitle                   

\section{Introduction}

The periodically driven planar pendulum is one  of the
simplest mechanical
systems which can exhibit chaotic behaviour. It is usually modelled as a
mathematical pendulum with linear damping and external or parametric
excitation \cite{Baker1990,Steindl1991}. 
Since the dynamics of various  systems of various kind  in  mechanical 
\cite{Baker1990,Steindl1991} or
electrical \cite{Cicogna1987,Strogatz1994} engineering  
can be described by the same or the same type of differential equations as
the one for the driven pendulum  it is not a great surprise,
that many  methods invented to calculate
and control chaotic vibrations of nonlinear systems
have been tested  on this system where the mathematical parameters
appearing in the above mentioned equations can be given a well
interpretable physical meaning.

In this note  we focus on the planar pendulum with a vertically periodically
moving suspension point with an additional constant torque \cite{Andronov1966,Coullet2005}. 
The corresponding differential equation has the following form:
\begin{equation}
\label{eq1}
ml^2\ddot{\phi} + k \dot{\phi}
+m (lg - l\ddot{x}) \sin \phi = T, \hspace{1cm} x=a\sin \Omega t,
\end{equation}
where $m$ and $l$ denote the point mass and the length of the mathematical pendulum and
$T$ is the extra constant torque applied at the suspension point. $a$ is the amplitude of  
the vertical excitation $x(t)$ of the suspension point 
and $\Omega$ the corresponding frequency of excitation.
  
Introducing dimensionless time $\tau=\omega t$, Eq. \ref{eq1} in dimensionless 
form is
\begin{equation}
\label{eq2}
\ddot{\phi} + \alpha \dot{\phi}
+ (1 +\gamma\Omega'^2 \cos\Omega'\tau) \sin \phi =t_0,
\end{equation}
where the corresponding constants are given by:
$
\omega=\sqrt{g/l}, \quad
\Omega'=\Omega/\omega, \quad
\alpha=k/(ml^2\omega), \quad
\gamma=a/l, \quad
t_0=T_0/(ml^2\omega^2).
$

Our goal is to investigate the dynamics of Eq. \ref{eq2} concerning
 the occurrence of chaotic vibrations using different 
approximations.

\section{Melnikov's analysis}

In the first approximation, we assume that damping,
excitation and extra torque  are all small terms which we scale by
introducing a small parameter $\epsilon$ to Eq. \ref{eq2} 
\begin{equation}
\label{eq3}
\ddot{\phi} + \epsilon \tilde{\alpha} \dot{\phi}
+(1 +\epsilon \tilde{\gamma}\Omega'^2 \cos{\Omega'\tau}) \sin \phi =
\epsilon \tilde{t}_0,
\end{equation}
where $\tilde \alpha \epsilon = \alpha$, $\tilde \gamma \epsilon = \gamma$,
$\tilde t_0 \epsilon = 
t_0$, respectively. This allows us to apply
Melnikov's method \cite{Melnikov1963,Guckenheimer1983}  for the 
unperturbed symmetric potential \cite{Steindl1991}. The critical 
value of the excitation amplitude 
$\gamma_c$ which is given by a simple zero of Melnikov function  
\cite{Melnikov1963,Guckenheimer1983}:
\begin{equation}
\label{eq4}
\gamma_c=\frac{4}{\pi \Omega'^4} \left| -\alpha
+\frac{t_0 \pi}{8} \right| \sinh \left( \Omega'\frac{\pi}{2} \right).
\end{equation}
is (for $t_0=0$) very similar to that obtained in \cite{Steindl1991}, 
with $\gamma'_c$, differing by a constant coefficient depending on the torque $t_0$: 
\begin{equation}
\label{eq5}
\gamma_c=\gamma'_c \frac{\alpha}{ \left| -\alpha
+\frac{t_0 \pi}{8} \right|}. 
\end{equation}
\begin{vchfigure}[htb]
  \includegraphics[width=.5\textwidth,width=4.5cm,angle=-90]{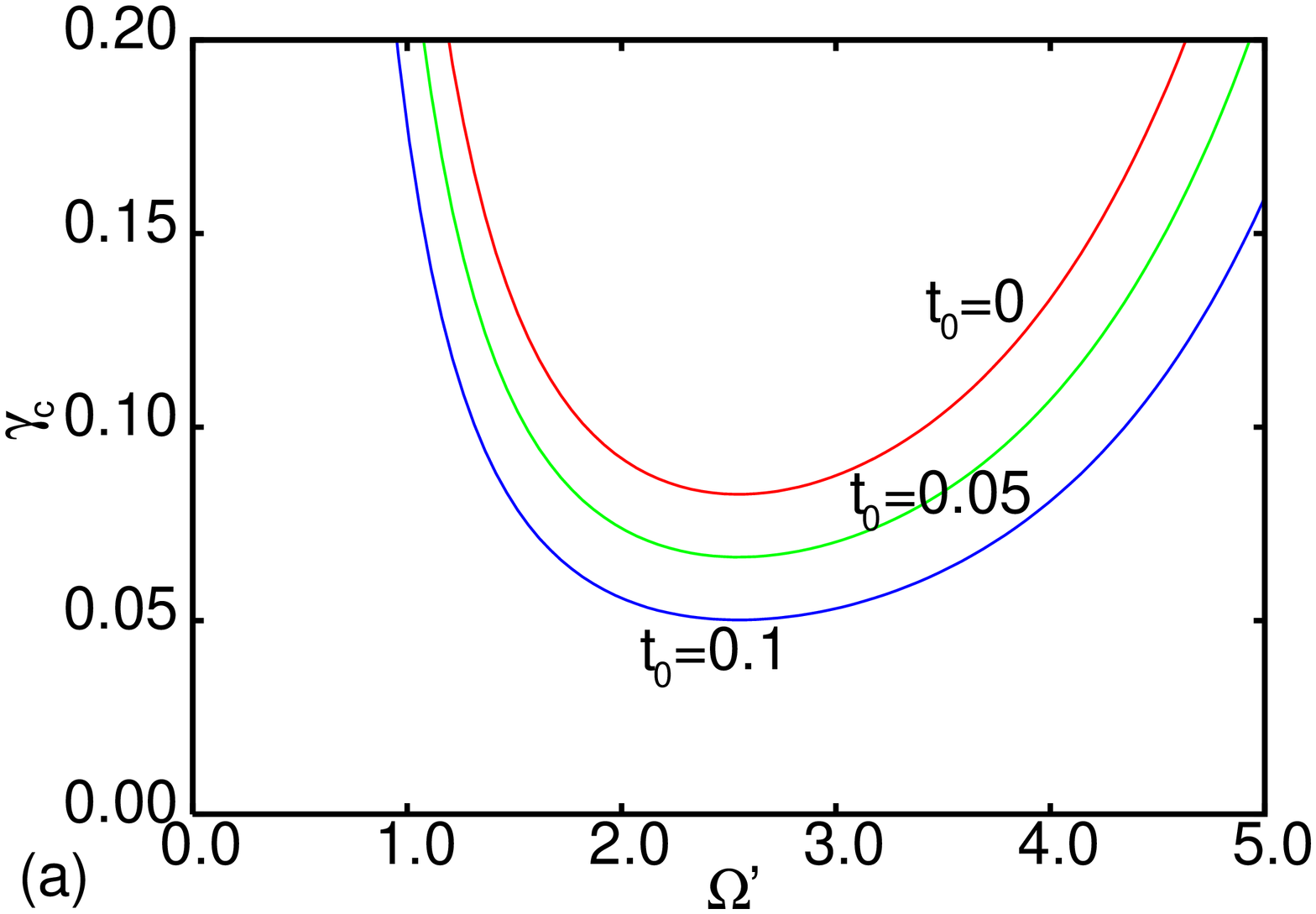}
  \includegraphics[width=.5\textwidth,width=4.5cm,angle=-90]{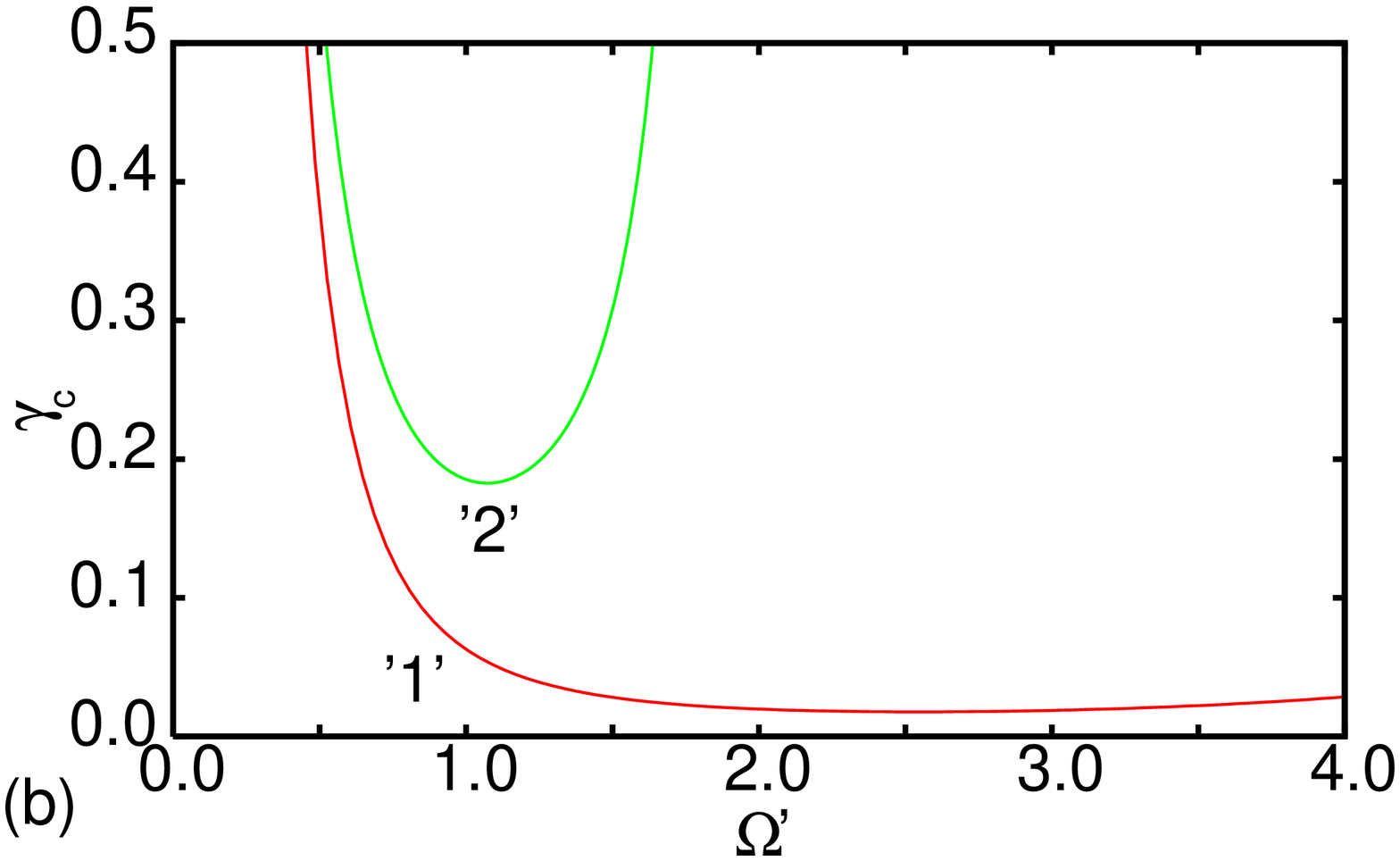}
\vchcaption{ \label{fig1} (a) Dependence of critical value of the excitation amplitude
$\gamma_c$ versus $\Omega'$ for $\alpha=0.1$ and three values
of a constant torque $t_0=0.0$, 0.05 and 0.1 in the case of heteroclinic transition.  (b) the same 
dependences 
for $t_0=0.2$ using two different approximations with  heteroclinic - '1' and homoclinic - 
'2' transitions.
Below the curves the system is regular while above transit to chaotic oscillations occurs.
Note, different scales in Figs. 1a and b.}
\end{vchfigure}

\begin{vchfigure}[htb]
\vspace{-0.5cm}
  \includegraphics[width=.5\textwidth,width=4.5cm,angle=-90]{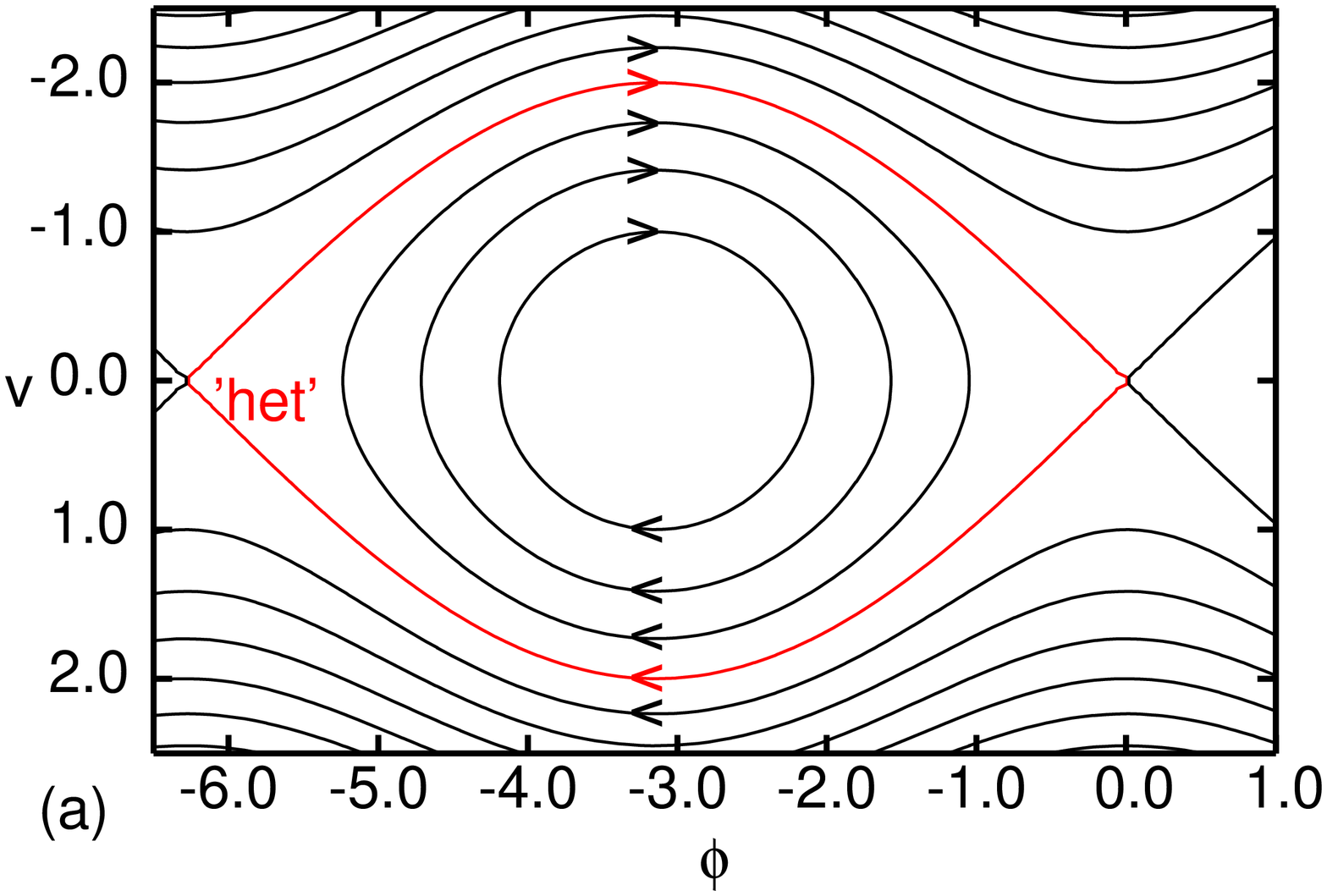}
  \includegraphics[width=.5\textwidth,width=4.5cm,angle=-90]{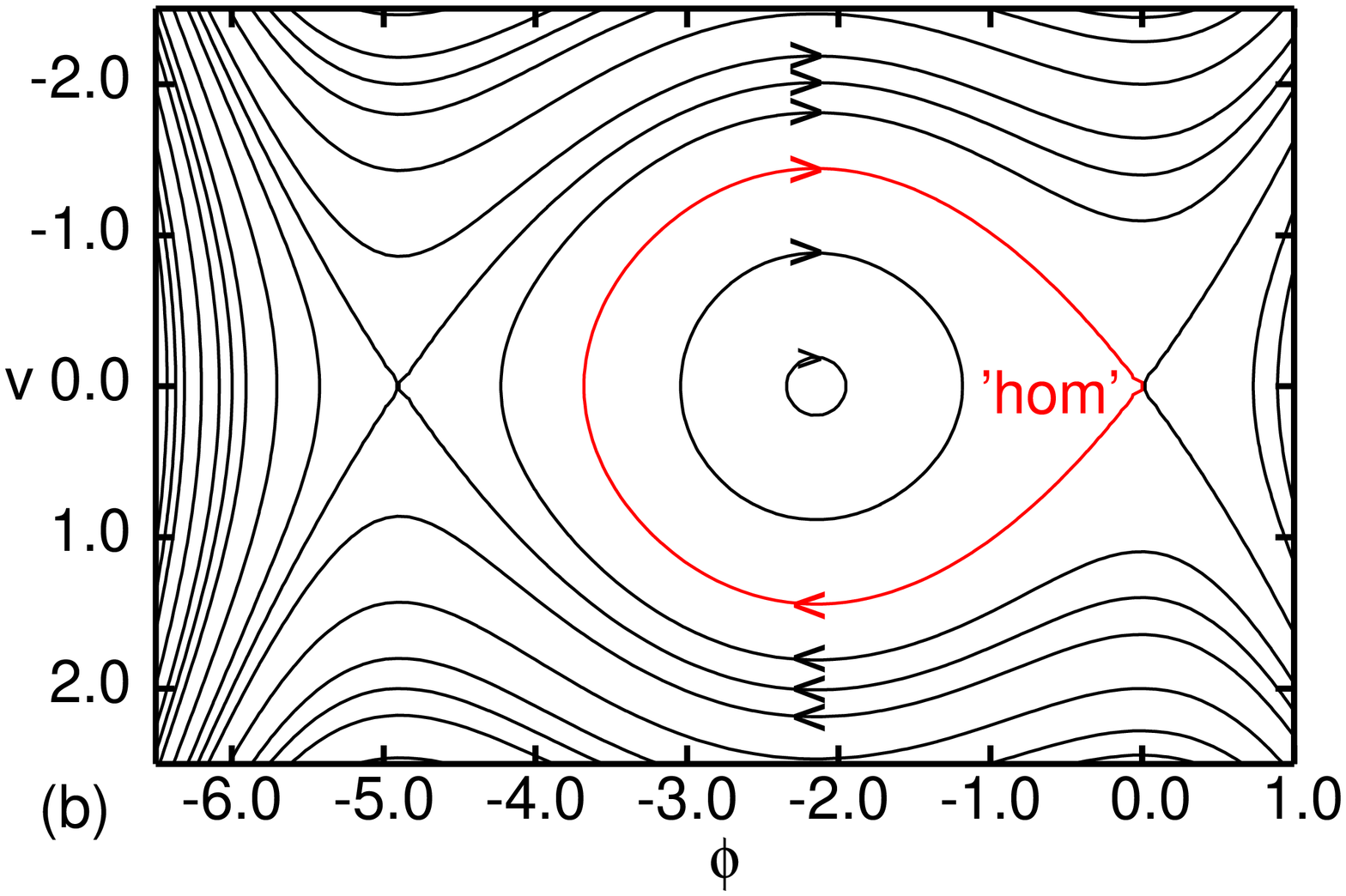}
\vchcaption{Phase trajectories (where $v=\dot \phi$)   for  $t_0=0.2$ and two 
different approximations,  with  
heteroclinic (a) and homoclinic (b) orbits (marked by 'het' and 'hom'), respectively.}
\end{vchfigure}
 In Fig. \ref{fig1}a $\gamma_c$ versus $\Omega'$ is plotted, showing 
the influence of the torque on the existence of chaotic dynamics for
(Eq. \ref{eq3}).
Note that these results are obtained for small torque compared to damping
$t_0 /\pi 8 \ll \alpha$. 

In the second approximation we treat the case of larger torque that is 
$ t_0 /(\pi 8) \sim \alpha $.
Here we use the recently developed method for 
non-symmetric potentials \cite{Litak2006a}. This approach deals with the 
equation 
\begin{equation}
\label{eq6}
\ddot{\phi} + \epsilon \tilde{\alpha} \dot{\phi}
+(1 +\epsilon \tilde{\gamma}\Omega'^2 \cos{\Omega'\tau}) \sin \phi -t_0 \cong
\ddot{\phi} + \epsilon \tilde{\alpha} \dot{\phi}
+\left[1 -t_0+\phi -\phi^3/6 \right]
+ \epsilon \tilde{\gamma}\Omega'^2 \cos{\Omega'\tau} \sin \phi
=0,
\end{equation}
where contrary to the first approximation $t_0$ is not taken to 
perturbations terms (contrary to Eq. \ref{eq3}) but
included into the 
effective unperturbed potential.
Consequently it breakes the potential symmetry and instead of a heteroclinc 
transition with two saddle 
points, a  homoclinic transition with a single saddle point is obtained 
\cite{Andronov1966} is obtained (see Fig. 2).

In Fig. \ref{fig1}b we compare results of both 
approximations.
Curve '1' corresponds to  the small torque approach with heteroclinc transition while '2' 
represents the results with $t_0$ included in  
the effective potential (a homoclinic transition).
Note, for  small 
frequencies both curves match  quite well but for higher 
frequencies there is a growing discrepancy between the two approximations. 




\end{document}